	\documentclass[10pt, draftclsnofoot, twocolumn]{IEEEtran}
	\usepackage{pifont,bm,multicol,amsfonts,amsmath,color,amssymb,graphicx, epsfig,cite,psfrag,subfigure,algorithm,enumerate,stfloats,algorithmic,epstopdf,balance}
    \usepackage{threeparttable}
	\usepackage{multirow}
    \usepackage[normalem]{ulem}
    \useunder{\uline}{\ul}{}

\usepackage{geometry}

\begin{document}
\vspace{-5mm}
\setlength{\abovecaptionskip}{0.235cm}
\title{{Graph Neural Network based Channel Tracking for Massive MIMO Networks}\vspace{-1mm}}
\author{Yindi Yang,  Shun Zhang,{ \emph{Member, IEEE,}} Feifei Gao, \emph{Fellow, IEEE,} \\ Jianpeng Ma,{ \emph{Member, IEEE,}}
Octavia A. Dobre, \emph{Fellow, IEEE}\vspace{-2mm}

    \thanks{This work of Y. Yang, S. Zhang and J. Ma is supported by the National Key  R$\&$D  Program of China under
Grant 2017YFB1010002 and the National Natural Science Foundation of
China under Grant (61931017, 61871455, 61901329), also supported in part
by the SAIC Science and Technology Foundation (No. 1911). The work of O. A. Dobre is supported by the Natural Sciences and Engineering Research Council of Canada (NSERC), through its Discovery program. (\emph{Corresponding anthor: Shun Zhang.})}

    \thanks{
    Y. Yang, S. Zhang and J. Ma are with the State Key Laboratory of Integrated Services Networks, Xidian University, Xi¡¯an 710071, P. R. China (Email: $\text{ydyangdu@163.com}$; zhangshunsdu@xidian.edu.cn; jpmaxdu@gmail.com).}

    \thanks{F. Gao is with  Institute for Artificial Intelligence, Tsinghua University
(THUAI), State Key Lab of Intelligent Technologies and Systems, Tsinghua
University, Beijing National Research Center for Information Science and
Technology (BNRist), Department of Automation,Tsinghua University Beijing,
P.R. China (email: feifeigao@ieee.org).}

    \thanks{{
    O. A. Dobre  is with Memorial University,
		St. John's, NL
		A1B 3X5, Canada (email: odobre@mun.ca). }}
}
\maketitle
\begin{abstract}
In this paper, we resort to {the} graph neural network (GNN) and
propose {the new channel} tracking
method for the massive multiple-input multiple-output
 networks under the high mobility scenario.
 We first utilize a small number of pilots  to achieve the initial channel estimation.
 Then, we represent the obtained channel data in the form of  {graphs} and describe the channel spatial correlation by the weights along the edges of the graph. Furthermore, we
 introduce the computation steps of the main unit for the GNN
and design {a} GNN-based channel tracking framework,
 which includes an encoder, a core network and a decoder.
{Simulation} results corroborate that our proposed GNN-based scheme can achieve better performance
than the works with {feedforward neural network}.

\end{abstract}

\maketitle
\thispagestyle{empty}



%
\IEEEpeerreviewmaketitle
\section{Introduction}
Massive {multiple-input multiple-output (MIMO)} {can} significantly improve the spectral and energy efficiencies,
{and} has become a key technology {for} 5G, where data traffic has increased dramatically\cite{efficiency2}.
{As is well known,}
obtaining {the} accurate channel state information (CSI) is {of great} importance in guaranteeing the performance of the massive MIMO {systems}\cite{MaTWC},
especially {under the high mobility scenario.}
In\cite{8410591}, {Ma \textit{et al.}} proposed {a} sparse Bayesian {learning-based} channel estimation algorithm for   time-varying massive MIMO {networks.}
{In\cite{7996914}, the authors designed a  channel tracking method based on spatial-temporal basis expansion model   under both time-varying and spatial-varying circumstances.}
{In \cite{8798669}, Han \textit{et al.}
fully exploited the delay and angular
reciprocity between the uplink and   downlink
to recover the time-varying downlink massive MIMO channels.}


{However, all these works \cite{8410591}-\cite{8798669} are
closely dependent on hypothetical statistical models. In the actual communication scenario,
the radio scattering conditions change rapidly with time, which may cause serious mismatch with the adopted mathematical model.}
{Deep learning (DL), aiming to {achieve a} performance gain from the data,
has undergone a renaissance with excellent performance and low complexity.}
{Hence, DL has been adopted to implement the signal processing tasks along the wireless radio links
 and has achieved superior performance.} {In \cite{A}, Al-Baidhani \textit{et al.}  used a deep autoencoder to estimate the received signal.}
 In \cite{Maxiaoli}, Ma \textit{et al.} developed a DL-based channel estimator for time-varying channels.
 In \cite{8322184}, {Wen \textit{et al.}} proposed a DL-based scheme to realize the downlink CSI sensing  to improve the quality of CSI reconstruction in  frequency division duplexing (FDD).
{Yang \textit{et al.} also applied DL to the doubly selective fading channel tracking in \cite{8672767}.}
{In \cite{CChun}, Chun \textit{et al.}  utilized the DL technique to implement  a  joint pilot design and channel estimation  for {multiuser MIMO channels}.}
{All these works utilize {either a feedforward neural network (FNN) or convolutional neural network, and basically implement the end-to-end learning through a black box operation.} {Thus, they cannot} clearly interpret the space correlation hidden in the data set.}


{Since graph neural network (GNN)} could effectively extract spatial relationships in data,
it has attracted many researchers' attention \cite{1806.01261}.
GNN merges the traditional model-based operation with the end-to-end learning,
and therefore is able to accurately capture the data features.
In fact, GNN has shown good performance in many fields, such as traffic prediction\cite{8917706} and medical diagnosis\cite{NIPS2017_7231}.
{For massive MIMO, the characterization of
the spatial correlation is vital to the low-complex channel tracking scheme design.
 Under the DL framework, the precise extraction of the spatial correlation would help the neural network to track the time-varying massive MIMO channels.}

%
In this paper, we propose an efficient online CSI prediction scheme based on GNN for the massive MIMO time-varying channels.  Firstly, we achieve the initial CSI with the traditional least square (LS) estimation.
Then, we characterize the achieved CSI with the graph data structure and extract
the channel spatial information from the edges of the graph.
Finally, we present
the main computation steps of GNN and construct
the GNN-based channel tracking framework. In order to fully capture the channel time correlation,
{we combine the time adjacent graphs for the initial CSI  into one graph, and feed it into the tracking framework}.
\vspace{-2mm}
\section{ System Model}
%
We consider {a}  massive MIMO system, which contains one base station (BS) and one user.
{BS is equipped with $N_r$ antennas in the
form of uniform linear array, and the user {is} equipped with single antenna.}
\vspace{-2mm}
\subsection{Time-varying Channel Model}
Due to the Doppler shift caused by the user's motion, the channel between BS and user is assumed to be time-varying.
{Correspondingly, the {uplink} channel at time $n$ can be written as}
{\begin{align}\label{channel model}
\mathbf{h}(n)=\sum\limits_{i=1}^{N_p}\alpha_{i} e^{j2\pi n\nu_iT_s}\mathbf{a}({\theta}_{i}){,}
\end{align}}where {$\mathbf h(n)=[h_1(n),h_2(n),\ldots,h_{N_r}(n)]^T$,} {with {$(\cdot)^{T}$ as the transpose} operator}, $\alpha_i\sim \mathcal{CN}(0,\sigma_\alpha^2)$ {denotes}
the propagation gain along the {$i$-th} path with {the average power} $\sigma_\alpha^2$,
$\nu_i$ is the Dropper shift for the {$i$-th} path, and $T_s$, $N_p$ separately represent the system sampling period and the number of scattering paths.
Moreover, the spatial steering vector  $\mathbf{a}(\theta_i)$ is defined as
\begin{align}
\mathbf{a}(\theta_{i})=\left[1,e^{-j\frac{2\pi d}{\lambda}sin\theta_{i}},...,e^{-j\frac{2\pi d}{\lambda}(N_r-1)sin\theta_{i}}\right]^{T}
\end{align}
where $d$ is the distance between the adjacent antennas, $\lambda$ is the signal carrier wavelength, and $\theta_{i}$ denotes the direction of arrival of the {$i$-th path}.
\begin{figure}[!t]
\setlength{\abovecaptionskip}{0.cm}
\setlength{\belowcaptionskip}{-0.cm}
	\centering
	\includegraphics[width=3.3in]{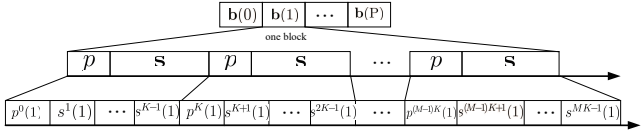}
	\caption{Transmitted signal structure.}
	\label{symbol}
\end{figure}
\vspace{-6mm}
\subsection{The Transmitted and Received Signal}
\subsubsection{Transmitted Signal Structure}The transmitted signal structure is shown in Fig. \ref{symbol}. Each frame contains $P$ blocks as $[\mathbf{b}(0),\mathbf{b}(1),...,\mathbf{b}(P)]$. Each block contains both $M$ groups of unknown {data} symbols $\mathbf{s}$ and $M$ pilot symbols denoted as $p$, where $\mathbf{s}\in \mathbb{C}^{(K-1)\times 1}$. We assume that a pilot signal $p$ is inserted between two adjacent groups of unknown signals $\mathbf{s}$.
\subsubsection{Received Signal}
The received signal {at the BS} at time $n$ is {expressed as}
\begin{align}
\mathbf{y}(n)=\mathbf{h}(n)x(n)+\mathbf{w}(n),
\end{align}
where $x(n)$ is the signal of user at time $n$ including pilots and  data signals, and $\mathbf{w}(n) \in \mathbb{C}^{N_r\times 1}$ denotes the additive white Gaussian noise (AWGN) with zero mean and {variance $\sigma_n^2$}.
\vspace{-2mm}
\subsection{Initial Channel Estimation}
Initially, we estimate the channels  at the pilots through the traditional LS estimator. For simplicity, {we  consider that the initial channel estimation based on each pilot
equals that of  the next $K-1$ signals of the corresponding pilot.}
{{If} the received signal at time {$n_p$} is a pilot, the LS estimation {corresponding to the} pilot and the next unknown signal {positions} can be {separately} expressed as:
 \begin{align}
 &\mathbf{h}^{\mathrm{LS}}(n_p)=\mathbf{y}(n_p)/x(n_p)\\ \notag
 &[\mathbf{h}^{\mathrm{LS}}(n_p\!+\!1),\!\ldots\!,\mathbf{h}^{\mathrm{LS}}(n_p\!+\!K\!-\!1)]\!=\![\mathbf{h}^{\mathrm{LS}}(n_p),\!\ldots\!,\mathbf{h}^{\mathrm{LS}}(n_p)],
 \end{align}
where $x(n_p)$ denotes the pilot at time $n_p$, {$\mathbf{h}^{\mathrm{LS}}(n_p)$} is the {estimated channels} at time $n_p$ obtained by LS estimation, and $[\mathbf{h}^{\mathrm{LS}}(n_p\!+\!1),\ldots,\mathbf{h}^{\mathrm{LS}}(n_p\!+\!K\!-\!1)]$  represents the LS {estimated} channels {for} $K-1$ unknown signals.} We uniformly represent the {initial estimation of the channels at time $n$ as $\mathbf{h}^{\mathrm{LS}}(n)=[h_1^{\mathrm{LS}}(n),h_2^{\mathrm{LS}}(n),\ldots,h_{N_r}^{\mathrm{LS}}(n)]^T$.}
\vspace{-2mm}
\section{{GNN-Based} Massive MIMO Channel Tracking}

\begin{figure}[!t]
\setlength{\abovecaptionskip}{0.cm}
\setlength{\belowcaptionskip}{-0.cm}
	\centering
	\includegraphics[width=3.7in]{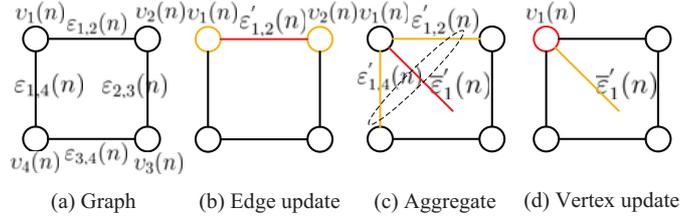}
	\caption{Updates in a GN block. Red indicates the element that is being updated, and orange indicates other elements which are involved in the update.}
	\label{graph}
\end{figure}
\subsection{Graph-based Massive MIMO Channel Representation}
\begin{figure*}[!t]
\setlength{\abovecaptionskip}{0.cm}
\setlength{\belowcaptionskip}{-0.cm}
	\centering \includegraphics[height=1.5in,width=5.1in]{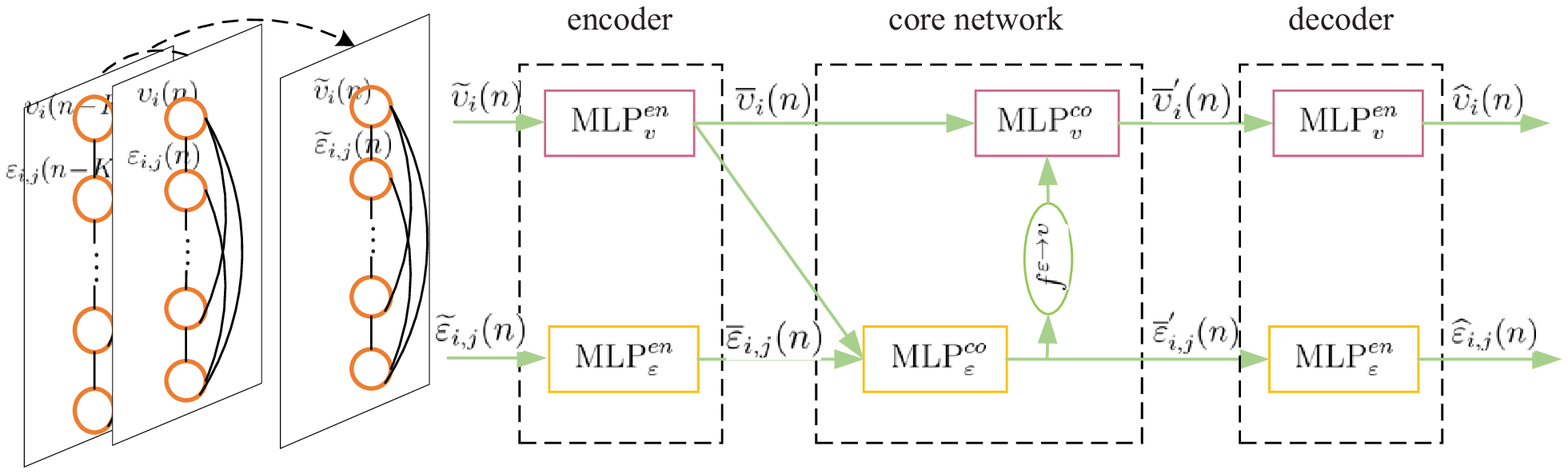}
	\caption{{The architecture of GNN-based massive MIMO channel tracking scheme.}}
	\label{system architecture}
\end{figure*}

The input of GNN is a dataset based on a graph. Thus, we should construct {a} graph ${\mathcal G}(n)=\{\mathcal V(n),\mathcal E(n)\}$ for $\mathbf h^{\mathrm{LS}}(n)$, where $\mathcal{V}(n)$ and $\mathcal{E}(n)$ are the vertex and edge sets, respectively.
In order to {simplify the neural network and speed up the network convergence,}
we treat each element in $\mathbf h^{\mathrm{LS}}(n)$ as one vertex.
{Moreover, the real and imaginary parts of {$h_i^{\mathrm{LS}}(n)$} are seen as two features of one vertex.}
Therefore, the element in $\mathcal V(n)$ can be expressed as
\begin{align}\label{nodes}
{\bm\upsilon_i(n)}=[\Re({h}^{\mathrm{LS}}_i(n)),\Im({h}^{\mathrm{LS}}_i(n))]^{T}{,}
 \end{align}
 where $\Re({h}^{\mathrm{LS}}_i(n))$ and $\Im({h}^{\mathrm{LS}}_i(n))$ are the real and imaginary parts of ${h}^{\mathrm{LS}}_i(n)$, respectively.
%
Then, we have $\bm\upsilon(n)=\left\{\bm\upsilon_1(n),\bm\upsilon_2(n),\ldots,\bm\upsilon_{N_r}(n)\right\}$.

To describe the spatial correlation between $h^{\mathrm{LS}}_{i}(n)$ and $h^{\mathrm{LS}}_{j}(n)$, we  define
the edge $\bm\varepsilon_{i,j}(n)$ between $\bm\upsilon_{i}(n)$ and $\bm\upsilon_{j}(n)$ as
\begin{small}
\begin{align}
{\bm\varepsilon_{i,j}(n)}=&\Bigg[\frac{\mathbb E\{\Re\{h^{\mathrm{LS}}_{i}(n)\}\Re\{h^{\mathrm{LS}}_{j}(n)\}\}}{\sqrt{\mathbb E\{\Re^2\{h^{\mathrm{LS}}_{i}(n)\}\}
\mathbb E\{\Re^2\{h^{\mathrm{LS}}_{j}(n)\}\}}},\notag\\
&\frac{\mathbb E\{\Im\{h^{\mathrm{LS}}_{i}(n)\}\Im\{h^{\mathrm{LS}}_{j}(n)\}\}}{\sqrt{\mathbb E\{\Im^2\{h^{\mathrm{LS}}_{i}(n)\}\}
\mathbb E\{\Im^2\{h^{\mathrm{LS}}_{j}(n)\}\}}}\Bigg]^T,
\end{align}
\end{small}where $\mathbb E(\cdot)$ denotes the expectation {operator}, $i, j=1,2,\ldots, N_r$, and
$i\neq j$.
Here, we utilize {$L$ adjacent samples $\mathbf h^{\mathrm{LS}}(n)$, $\mathbf h^{\mathrm{LS}}(n-1)$, \ldots, $\mathbf h^{\mathrm{LS}}(n-L+1)$}
to approximate $\mathbf e_{i,j}(n)$.
Before proceeding, we define the matrix
$\mathbf H^{\mathrm{LS}}_L(n)=[\mathbf h^{\mathrm{LS}}(n),\mathbf h^{\mathrm{LS}}(n-1),\ldots,\mathbf h^{\mathrm{LS}}(n-L+1)]$, and the matrix
$\mathbf{\bar H}^{\mathrm{LS}}_L(n)=\frac{1}{L}\sum\limits_{k=0}^{L-1}\mathbf h^{\mathrm{LS}}(n-k)\mathbf 1_{L}^T$. Then, the first element of
$\bm\varepsilon_{i,j}(n)$ is written as
\begin{small}
\begin{align}
&[\bm\varepsilon_{i,j}(n)]_1=\notag\\
&\frac{(\Re\{[\mathbf H^{\mathrm{LS}}_L(n)]_{i,:}-[\mathbf {\bar H}^{\mathrm{LS}}_L(n)]_{i,:}\})(\Re\{[\mathbf H^{\mathrm{LS}}_L(n)]_{j,:}-[\mathbf {\bar H}^{\mathrm{LS}}_L(n)]_{j,:}\})^T}{
\|\Re\{[\mathbf H^{\mathrm{LS}}_L(n)]_{i,:}-[\mathbf {\bar H}^{\mathrm{LS}}_L(n)]_{i,:}\}\|
\|\Re\{[\mathbf H^{\mathrm{LS}}_L(n)]_{j,:}-[\mathbf {\bar H}^{\mathrm{LS}}_L(n)]_{j,:}\}\|},\label{eq:corr}
\end{align}
\end{small}where $\|\mathbf a\|$ is the {$L_2$-norm} of  vector $\mathbf a$. Similarly, we
can evaluate the second entry of $\bm \varepsilon_{i,j}(n)$.

\vspace{-2mm}
\subsection{The Computation Steps of {Graph Network (GN)}}
{The main unit of the GNN framework is the GN block, whose input and output are graphs constructed from sample data.
Once $\mathcal G(n)$ flows into the GN block, the computation is sequentially performed from edge to node.
Specifically, this process includes three sub-functions, {namely} the edge updating unit $f^\varepsilon$, the node updating unit $f^v$, and the aggregating unit $f^{\varepsilon\rightarrow \upsilon}$}, as shown in Fig. \ref{graph}.
The input of each $f^\varepsilon$ includes $\bm\varepsilon_{i,j}(n)$ and its connection vertices $\bm\upsilon_{i}(n)$ and $\bm\upsilon_{j}(n)$, and can be implemented by the NN as
\begin{align}
{\bm\varepsilon_{i,j}^{'}(n)}& \leftarrow f^\varepsilon(\bm\varepsilon_{i,j}(n),\bm\upsilon_{i}(n),\bm\upsilon_{j}(n))\\ \notag
&=\mathrm{NN}_\varepsilon(\bm\varepsilon_{i,j}(n),\bm\upsilon_{i}(n),\bm\upsilon_{j}(n)),
\end{align}
where $\bm\varepsilon_{i,j}^{'}(n)$ is the updated edge of $\bm\varepsilon_{i,j}(n)$
 and $\mathrm{NN}_\varepsilon$ is $f^\varepsilon$'s corresponding NN.

Correspondingly, $f^{\varepsilon\rightarrow \upsilon}$ collects all the updated edges, which are connected with
$\bm\upsilon_i(n)$, into the aggregating edge ${\bm\varepsilon}_{i}^{'}(n)$ as
\begin{align}\label{aggregate}
{{\bm\varepsilon}_{i}^{'}(n)}&\leftarrow f^{\varepsilon\rightarrow \upsilon}(\{{\bm\varepsilon}_{i,j}^{'}(n)\}_{{\bm\upsilon_j\in\mathcal{N}(\bm\upsilon_i(n))}})\\ \notag
&=\sum_{\bm\upsilon_j\in\mathcal{N}(\bm\upsilon_i(n))}{\bm\varepsilon}_{i,j}^{'}(n),
\end{align}where $\mathcal{N}(\bm\upsilon_i)$ denotes the neighbor vertices of  $\bm\upsilon_i$ and $\overline{\bm\varepsilon}^{'}(n)$ is the aggregate edge.

With ${\bm\varepsilon}_{i}^{'}(n)$, $f^\upsilon$ would renew $ \bm\upsilon_i(n)$ as
\begin{align}
{\bm\upsilon_i^{'}(n)}\leftarrow f^\upsilon({\bm\varepsilon}_i^{'}(n),\bm\upsilon_i(n))=\mathrm{NN}_v({\bm\varepsilon}_i^{'}(n),\bm\upsilon_i(n)),
\end{align}
where $\mathrm{NN}_\upsilon$ is the NN for $f^\upsilon$ {and} $\bm\upsilon_i^{'}(n)$ denotes the updated vertex of $\bm\upsilon_i(n)$.
For clarity, we present the detailed steps of GN in Algorithm 1.
\renewcommand\arraystretch{1.0}
\begin{algorithm}[h]
\caption{The computation steps of  GN}
\begin{algorithmic}[1]
\REQUIRE $\mathcal{G}(n)=\Big\{\mathcal{E}(n),\mathcal{V}(n)\Big\}$
\FOR{each edge $\bm\varepsilon_{i,j}$, $i,j=1,...,N_r$}
\STATE Compute edge-wise features, \\$\bm\varepsilon_{i,j}^{'}(n)\leftarrow f^\varepsilon(\bm\varepsilon_{i,j}(n),\bm\upsilon_{i}(n),\bm\upsilon_{j}(n))$
\ENDFOR
\FOR{each vertex $\bm\upsilon_i(n)$, $i = 1,...,N_r$}
\STATE Calculate the aggregating edge \\$\bm\varepsilon_{i}^{'}(n)\leftarrow f^{\varepsilon\rightarrow \upsilon}(\left\{{\bm\varepsilon}_{i,j}(n)\right\}_{{j=\mathcal{N}(i)}})$\\
\STATE Obtain the vertex-wise features \\$\bm\upsilon_{i}^{'}(n)\leftarrow f^{\upsilon}(\bm\varepsilon^{'}_{i}(n),\bm\upsilon_{i}(n))$
\label{code:TrainBase:pos}
\ENDFOR
\STATE Achieve the updated graph $\mathcal G^{'}(n)\!=\!\{\mathcal V^{'}(n),\mathcal E^{'}(n)\}$, where ${\mathcal{V}^{'}}(n)\!=\!\left\{\bm\upsilon_i^{'}(n)\right\}_{i=1:N_{r}}$,
$\mathcal{E}^{'}(n)=\left\{{\bm\varepsilon}_{i,j}^{'}(n)\right\}_{{i=1:N_r,j=\mathcal{N}(i)}}$
\RETURN   $\mathcal{{G}}^{'}(n)=\Big\{\mathcal{{E}}^{'}(n),\mathcal{{V}}^{'}(n)\Big\}$
\end{algorithmic}
\end{algorithm}
\vspace{-2mm}
\subsection{GNN-based Architecture for Channel Tracking}

In order to track the massive MIMO channels, we design the GNN-based framework in Fig. \ref{system architecture}.
 This architecture includes an encoder, a core network and a decoder.
The historical channel samples are fed into the encoder
{to initialize the core network,}
the core network uses the graph structure to update nodes and edges, and the decoder independently decodes the edge and vertex attributes. The output of the decoder is $\mathcal {\widehat G}(n)$.

To better achieve the time-correlation of the massive MIMO channels,
we combine the channel  graph $\mathcal { G}(n)$ at the current time with the graph $\mathcal { G}(n-K)$ {for the time $n-K$} to regenerate the graph $\widetilde{\mathcal { G}}(n)$ as the input of the encoder. {Correspondingly, $\widetilde{\mathcal { G}}(n)$} can be denoted as
\begin{align}
{\mathcal{ \tilde G}(n)}=\mathrm{concat}(\mathcal G(n),\mathcal {{G}}(n-K)),
\end{align}
and its vertices $\widetilde{\bm\upsilon}_i(n)$ and edges $\widetilde{\bm\varepsilon}_{i,j}(n)$ can be expressed as
\begin{align}
{\widetilde{\bm\upsilon}_i(n)}=&\left[\left(\bm\upsilon_i(n)\right)^{T},\left({\bm\upsilon}_i(n-K)\right)^{T}\right]^{T},\\ \notag
{\widetilde{\bm\varepsilon}_{i,j}(n)}=&\left[\left(\bm\varepsilon_{i,j}(n)\right)^{T},\left({\bm\varepsilon}_{i,j}(n-K)\right)^{T}\right]^{T},
\end{align}
where {${\bm\upsilon}_i(n-K)$, ${\bm\varepsilon}_{i,j}(n-K)$ separately
represent the vertex and the edge of $\mathcal{ G}(n-K)$.}

As well known, {the DL performance} is closely related with
the feature representation. Thus, in our structure, we utilize the encoder part
to extract and describe the latent features of $\mathcal{\tilde G}(n)$.
Specifically, different {multilayer perceptrons (MLPs)} are resorted to independently {extract} the features
 of the vertices and  edges in $\mathcal{ \tilde G}(n)$.
For $\widetilde{\bm\varepsilon}_{i,j}(n)$ and $\widetilde{\bm\upsilon}_{i}(n)$, the operations can be explicitly written as
\begin{align}
{\overline{\bm\varepsilon}_{i,j}(n)}\!=\!\mathrm{MLP}_\varepsilon^{en}(\widetilde{\bm\varepsilon}_{i,j}(n)),\overline{\bm\upsilon}_{i}(n)\!=
 \!\mathrm{MLP}_\upsilon^{en}(\widetilde{\bm\upsilon}_{i}(n)),
\end{align}
 where $\overline{\bm\upsilon}_{i}(n)$ and $\overline{\bm\varepsilon}_{i,j}(n)$ are the resultant features, while
 $\mathrm{MLP}_\varepsilon^{en}$ and $\mathrm{MLP}_\upsilon^{en}$ are the adopted MLPs during the encoder part.

 Then, the graph $\overline{\mathcal G}(n)$ formed by $\overline{\bm\upsilon}_{i}(n)$, $\overline{\bm\varepsilon}_{i,j}(n)$ {flows} into
 the core network, which implements Algorithm 1 to achieve updated graph $\overline{\mathcal G}^{'}(n)$. Different
 from the previous subsection, we utilize the MLP, i.e, $\mathrm{MLP}_\varepsilon^{co}$, to conduct
 $f^\varepsilon$, while $\mathrm{MLP}_\upsilon^{co}$ is used to fulfil  $f^\upsilon$. Then, within this part,
 the three sub-functions of GN algorithm can be reexpressed as
\begin{align}
&{\overline{\bm\varepsilon}_{i,j}^{'}(n)}=\mathrm{MLP}_\varepsilon^{co}(\overline{\bm\varepsilon}_{i,j}(n),\overline{\bm\upsilon}_{i}(n),\overline{\bm\upsilon}_{j}(n)), \notag \\
&{\overline{\bm\varepsilon}_{i}^{'}(n)}=\sum_{\overline{\bm\upsilon}_j\in\mathcal{N}(\overline{\bm\upsilon}_i(n))}{\overline{\bm\varepsilon}}_{i,j}^{'}(n),\notag \\
&{\overline{\bm\upsilon}_i^{'}(n)}=\mathrm{MLP}_\upsilon^{co}(\overline{\bm\upsilon}_i(n),\overline{\bm\varepsilon}_{i}^{'}(n)).
\end{align}

In the decoder, we recover  $\mathcal{\widehat G}(n)$ from $\mathcal {\bar{G}}^{'}(n)$, where
MLPs $\mathrm{MLP}_v^{de}$ and $\mathrm{MLP}_e^{de}$ are used for vertex $\overline{\bm\upsilon}_{i}^{'}(n)$  and edge $\overline{\bm\varepsilon}_{i,j}^{'}(n)$, respectively. Similarly, the process
can be defined as
\begin{align}
{\widehat{\bm\varepsilon}_{i,j}(n)}\!=\!\mathrm{MLP}_\varepsilon^{de}(\overline{\bm\varepsilon}_{i,j}^{'}(n)),
{\widehat{\bm\upsilon}_{i}(n)}\!=\!\mathrm{MLP}_\upsilon^{de}(\overline{\bm\upsilon}_{i}^{'}(n)).
\end{align}

Finally, we reorganize the prediction output graph $\mathcal{\widehat G}(n)$ composed of $\widehat{\bm\upsilon}_{i}(n)$ and $\widehat{\bm\varepsilon}_{i,j}(n)$ to obtain $\widehat{\mathbf h}(n)$ as
\begin{align}
\widehat{\mathbf h}(n)=
&\left[\left[\widehat{\bm\upsilon}_1(n)\right]_0+j\left[\widehat{\bm\upsilon}_1(n)\right]_1,\right. \\ \notag
&\left.\ldots,[\widehat{\bm\upsilon}_{N_r}(n)]_0+j[\widehat{\bm\upsilon}_{N_r}(n)]_1\right]^T{,}
\end{align}
where $[\widehat{\bm\upsilon}_i(n)]_0$ and $[\widehat{\bm\upsilon}_i(n)]_1$ represent the first and second elements of the $\widehat{\bm\upsilon}_i(n)$, respectively.



\vspace{-2mm}
\subsection{ Model Training and {Deployment}}
Our proposed channel scheme has two stages, i.e., the training and the deployment.
 In the first stage, we  utilize the off-line learning scheme to train the GNN-based architecture to minimize the error between $\widehat{\mathbf{h}}(n)$ and $\mathbf{h}(n)$.

\begin{figure}[!t]
\setlength{\abovecaptionskip}{0.cm}
\setlength{\belowcaptionskip}{-0.cm}
	\centering
	\includegraphics[height=1.6in,width=2.7in]{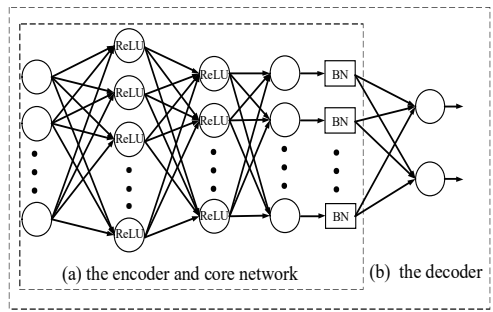}
	\caption{MLPs structure in the encoder, the core network{, and} the decoder. }
	\label{MLP}
\end{figure}

In the encoder and core network, for each MLP, we apply the same learning structure as shown in {Fig. \ref{MLP}(a).}
{Except for} the output layer, the rectified linear unit (ReLU) activation function is utilized for each neuron.
In addition, batch-normalization (BN) operation is utilized after the output layer to avoid gradient disappearance.
{The only difference between MLP in the encoder and that in the decoder is that the linear {fully-connect (FC)}  layer with two output neurons is placed at the output of BN.} The structure is shown in {Fig. \ref{MLP}(b).}

{Without loss of generality,} we use the {mean square error (MSE)} of the channel estimation as the loss function,
{and} add the {$L_2$-norm} as the regularization function to improve the generalization ability. Therefore,
the loss function can be written as
\begin{align}
\mathcal{L}(\bm{\xi})=\frac{1}{N_rM_{tr}}\sum_{\mu=1}^{M_{tr}}\parallel \mathbf{h}^{(\mu)}(n)-\mathbf{\widehat{h}}^{(\mu)}(n) \parallel^2
+\kappa\mathbf{\bm\xi}^\mathrm{T}\mathbf{\bm\xi},
\end{align}
where $M_{tr}$ is the batch size, $\bm{\xi}$ are the weight parameters of MLPs to be learned, and {$\kappa$} represents
the regularization coefficient.
The {adaptive moment estimation (ADAM) \cite{ADAM}} optimizer algorithm is adopted to achieve the optimal model parameters  as $\bm \xi^*$.

In the on-line deployment stage, we load the trained parameters $\bm{\xi}^*$, {pass} the input data with the same structure as the training stage, and
{track the massive MIMO channels}.

\vspace{-2mm}
\section{{Simulation Results}}
In this section, we numerically evaluate the performance of
our proposed {GNN-based massive MIMO channel tracking scheme.}
{{The number of antennas at BS is set as} $N_r$ =
32, and the channel attenuation {is complex Gaussian distributed as} $\alpha_i\sim\mathcal{CN}(0,1)$. {Moreover},  {the number of paths $N_p$ is set as} 20, the direction of arrival $\theta_i$ follows the uniform
distribution over [$-\pi$,$\pi$],} {the} sampling time {$T_s$} is $2\times 10^{-5}s$,  {the} carrier frequency is {3~GHz}, {and the antenna spacing $d$ equals to $\frac{\lambda}{2}$}.
\renewcommand\arraystretch{1.2}
\begin{table}[]
\setlength{\abovecaptionskip}{0.cm}
\setlength{\belowcaptionskip}{-0.1cm}
\caption{Default GNN parameters.}
\label{parameters}
\begin{tabular}{c|c|c|c}
\cline{1-4}
{Parameters}     &encoder        & core network   &decoder                     \\
\cline{1-4}
{Neurons in hidden layers}    & (16, 16)        &  (16, 16)   &  (16, 16, 8)    \\
\cline{1-4}
{Neurons in output layers}    & 8        &  8   &  2  \\
\cline{1-4}
{Exponential decay rates}    & \multicolumn{3}{c}{(0.9, 0.999)}   \\
\cline{1-4}
{Activation function}    & \multicolumn{3}{c}{ReLU}   \\
\cline{1-4}
{Batch size}    & \multicolumn{3}{c}{20}   \\
\cline{1-4}
\end{tabular}
\end{table}

The default parameters of the GNN-based estimator are given in TABLE \ref{parameters}.
In the encoder, the numbers of  neurons in the input layers   is 2 because each vertex and edge have two attributes.
The regularization coefficient {$\kappa$} is taken as 0.1 {to avoid overfitting}.
To illustrate the performance of the GNN, we compare it with FNN and convolutional neural network (CNN).  {The} number of each layer neurons in FNN is (64, 256, 128, 64).  {Moreover, in CNN, we form the massive MIMO channel vectors to the $N_r \times K$ data blocks over the space-time domain
and use 8-layer network structure. Correspondingly, each layer uses 64  convolution filters of size $3 \times 3$. The rest of the default parameters are the same as {that for} GNN.}
Moreover, the number of training samples is 10000.


{First, we provide TABLE II {to show} the  MSE versus $L$ at the user speed of 50 m/s, {when}  $K$ is 5.
As can be seen from TABLE II, the MSE decreases with  increasing  $L$ and  {quickly attains} its steady state.
In our scheme, the final results of the GNN output are determined by both the initial input
and the performance gain of the GNN. When $L$ is small, the initial correlation captured by (\ref{eq:corr}) is too coarse. When $L$ becomes large enough,
the available correlation from  (\ref{eq:corr}) suffices. Since the GNN-based estimator can achieve a good performance at $L=10$, we use this value in
the following simulations.}


When the signal-to-noise ratio (SNR) is 20 dB and the user's moving speed is 50 m/s, the corresponding {MSEs with respect to} different $K$ {values} and
 learning {rates}  are summarized in TABLE II. {Notice that} the best results {are} presented
in  bold font, and the learning rates represent the step size of ADAM algorithm for gradient learning. The performance of the GNN-based estimator degrades as the {number} of symbols {$K$} increases. In addition, we compare the performance of the {GNN-based estimator with the {FNN-based} one for different learning rates}.
As can be seen from the table, the performance of the {GNN-based estimator} is best when the learning rate is 0.001, and the MSE of the
{GNN-based estimator} is significantly lower than the {FNN-based one}.
\renewcommand\arraystretch{1.2}
\begin{table}[]
\setlength{\abovecaptionskip}{0.cm}
\setlength{\belowcaptionskip}{-0.2cm}
\caption{{MSE versus $L$}. }
\begin{tabular}{c|ccccc}
\cline{1-6}
\multirow{2}{*}{} & \multicolumn{5}{c}{{MSE}}                     \\
\cline{2-6}
{Method}     &{$L$=5}       & {$L$=10}       & {$L$=20}      & {$L$=30}  &{$L$=40}\\
\cline{1-6}
{GNN}        &  {0.0038}   &  {0.0035}   & {0.0034}   &{ 0.0035} &{0.0035}    \\
\cline{1-6}
\end{tabular}
\end{table}
\renewcommand\arraystretch{1.2}
\begin{table}[]
\setlength{\abovecaptionskip}{0.cm}
\setlength{\belowcaptionskip}{-0.2cm}
\caption{MSE versus K and learning rate. }
\begin{tabular}{c|c|cccc}
\cline{1-6}
\multirow{2}{*}{}   & \multirow{2}{*}{} & \multicolumn{4}{c}{MSE}                     \\
\cline{3-6}
        {Learning rate }                    &{Method}     & $K$=2       & $K$=5       & $K$=10      & $K$=15 \\
\cline{1-6}
\multirow{2}{*}{$1\!\times \!10^{-2}$}    & FNN        &  0.0177   &  0.0185   &  0.0199   & 0.0231    \\
                                          & GNN        &  0.0044   &  0.0048   &  0.0054   &   0.0057  \\
\cline{1-6}
\multirow{2}{*}{$1\!\times \!10^{-3}$}    & FNN        &  0.0067   &  0.0075   & 0.0075    &   0.0078   \\
                                          & GNN        &  $\mathbf {0.0030}$   &  $\mathbf{0.0035}$   & $\mathbf{0.0037}$    &   $\mathbf{0.0041}$  \\
\cline{1-6}
\multirow{2}{*}{$1\!\times \!10^{-4}$}    & FNN        &  0.0057   &  0.0061    &  0.0063   &   0.0065  \\
                                          & GNN        &  0.0035   &  0.0036   &  0.0040   &   0.0043 \\
\cline{1-6}
\end{tabular}
\end{table}
\begin{figure}[!t]
\setlength{\abovecaptionskip}{0.cm}
\setlength{\belowcaptionskip}{-0.cm}
	\centering
	\includegraphics[width=65mm]{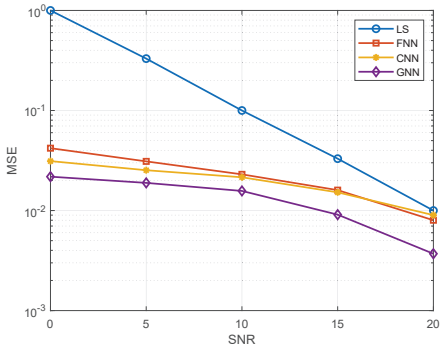}
	\caption{MSE {versus} SNR when user's motion speed is 50 m/s. }
	\label{SNR}
\end{figure}
{In Fig. \ref{SNR}, we compare the performance of  four estimators,
namely the GNN-based estimator, {FNN-based estimator},
CNN-based estimator and {LS estimator}, {for} different SNR.
$K$ for  GNN-based, CNN-based and FNN-based estimators is {set as}  10. The learning rates for GNN, CNN and FNN are $1\times 10^{-3}$, {$1\times 10^{-4}$, and $1\times 10^{-4}$ according to the TABLE III}, respectively.}
{From Fig. \ref{SNR}, we {make} the following observations.}
{As the SNR increases, the MSE of the four estimators  gradually decreases.
Among the above four estimators, the GNN-based estimator can achieve the best performance, especially {in the high} SNR region.}
\begin{figure}[!t]
\setlength{\abovecaptionskip}{0.cm}
\setlength{\belowcaptionskip}{-0.cm}
	\centering
	\includegraphics[width=65mm]{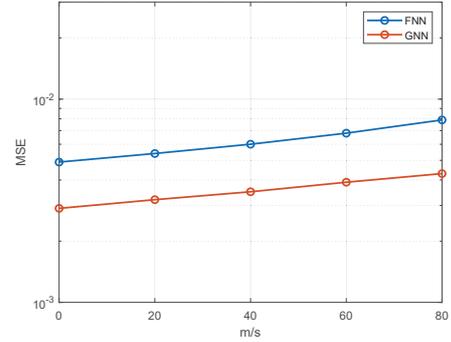}
	\caption{MSE {versus} user's motion speed {at} SNR=20 dB. }
	\label{Speed}
\end{figure}
The user's moving speed greatly affects the CSI of the time-varying channel and  the performance of the estimator. In Fig. 6, we show the channel estimation MSE {versus }the user's moving speed. We set the learning rates and $K$ {to} the same values as in {Fig. \ref{SNR}}. As the speed increases, the performance of both {FNN- and  GNN-based estimators} decreases{; however,} the MSE of the {latter} is always lower than {that of} the {former}. {In other words}, the GNN-based estimator {is} more applicable {under}  high mobility scenario.
\vspace{-2mm}
\section{Conclusion}
In this paper, we examined GNN-based massive MIMO channel tracking.
We fully exploited the data representation capability of the graph
to accurately characterize the channel spatial information.
A tracking framework with one encoder, one core network, and one decoder was constructed,
where the graph combination operation was resorted to capture the time correlation information of
the massive MIMO channels. The numerical experiments verified that our scheme
could achieve better performance than that with FNN under high mobility scenario.
\vspace{-5mm}

\end{document}